\documentclass[journal]{IEEEtran}

\ifCLASSINFOpdf
  \usepackage[pdftex]{graphicx}
  \graphicspath{{./figs/}}
  \DeclareGraphicsExtensions{.pdf,.jpeg,.png}
\else
\fi
\usepackage[cmex10]{amsmath}
\usepackage{mathtools}
\usepackage{amssymb}
\usepackage{algorithm}
\usepackage{algpseudocode}

\usepackage{array}
\PassOptionsToPackage{hyphens}{url}\usepackage{hyperref}

\usepackage{biograph}        
\usepackage{afterpage}
\usepackage{array}
\usepackage{booktabs}

\usepackage{cite}
\usepackage{multiro                                                                                                                                                                                     w} 
\usepackage{rotating}

\usepackage{float}
\usepackage{color}
\usepackage{soul}

\usepackage{colortbl}

\def\por1{\partial}

\newcolumntype{S}{>{\centering\arraybackslash} m{.4\linewidth} }

\makeatletter
  \newcommand\tinyv{\@setfontsize\tinyv{5pt}{7}}
\makeatother

\newlength{\hspacephantom}
\begin{document}
\DeclareGraphicsExtensions{.pdf,.jpeg,.png}

\title{A Note on Efficiency of Downsampling and Color Transformation in Image Quality Assessment}



\author{Hossein Ziaei Nafchi \textnormal{and} Mohamed Cheriet \\ Synchromedia Laboratory for Multimedia Communication in Telepresence \\
\'Ecole de technologie sup\'erieure, Montreal (QC), Canada H3C 1K3}

\markboth{}%
{}
%


\maketitle

\begin{abstract}
Several existing and successful full reference image quality assessment (IQA) models use linear color transformation and downsampling before measuring similarity or quality of images. This paper indicates to the right order of these two procedures and that the existing models have not chosen the more efficient approach. In addition, efficiency of these metrics is not compared in a fair basis in the literature. 
\end{abstract}


%
\IEEEpeerreviewmaketitle


\maketitle

\section{Introduction}
\label{sec:intro}


\IEEEPARstart{A}{pplications} of perceptual image quality assessment (IQA) in image and video processing, such as image acquisition, image compression, image restoration and multimedia communication, have led to the development of many IQA metrics. IQA models (IQAs) mimic the average quality predictions of human observers. This quality prediction is an easy task for the human visual system (HVS) and the result of the evaluation is reliable. Automatic quality assessment, e.g. objective evaluation, is not an easy task because images may suffer from various types and degrees of distortions. 

Among IQAs, the conventional metric mean squared error (MSE) and its variations are widely used in full-reference IQA (FR-IQA) applications because of their simplicity. However, in many situations, MSE does not correlate with the human perception of image fidelity and quality \cite{scope2008, spm2009}. Because of this limitation, a number of IQAs have been proposed to provide better correlation with the HVS \cite{SSIM, MSSSIM, MAD, FSIM, VSI, SCQI}. In general, these better performing models take into account the structural and/or color distortions.   

\begin{figure}[htb]
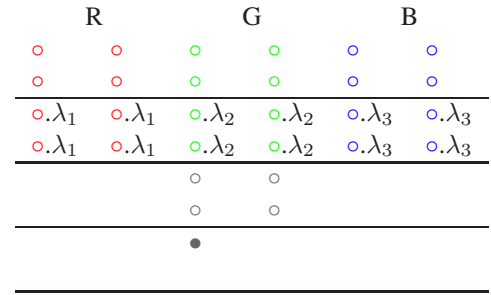

\centering
\begin{tabular}{llllll}
\multicolumn{2}{c}{R}                          & \multicolumn{2}{c}{G}                          & \multicolumn{2}{c}{B}                          \\
\color[rgb]{1,0,0}$\circ$                      & \multicolumn{1}{l}{\color[rgb]{1,0,0}$\circ$} & \color[rgb]{0,1,0}$\circ$                      & \multicolumn{1}{l}{\color[rgb]{0,1,0}$\circ$} & \color[rgb]{0,0,1}$\circ$                      & \multicolumn{1}{l}{\color[rgb]{0,0,1}$\circ$} \\
\color[rgb]{1,0,0}$\circ$                      & \multicolumn{1}{l}{\color[rgb]{1,0,0}$\circ$} & \color[rgb]{0,1,0}$\circ$                      & \multicolumn{1}{l}{\color[rgb]{0,1,0}$\circ$} & \color[rgb]{0,0,1}$\circ$ & \multicolumn{1}{l}{\color[rgb]{0,0,1}$\circ$} \\ \hline 
\multicolumn{1}{l}{{\color[rgb]{1,0,0}$\circ$}.$\lambda_1$}  & \multicolumn{1}{l}{{\color[rgb]{1,0,0}$\circ$}.$\lambda_1$}                       & \multicolumn{1}{l}{{\color[rgb]{0,1,0}$\circ$}.$\lambda_2$}  & \multicolumn{1}{l}{{\color[rgb]{0,1,0}$\circ$}.$\lambda_2$}                         & \multicolumn{1}{l}{{\color[rgb]{0,0,1}$\circ$}.$\lambda_3$}  & \multicolumn{1}{l}{{\color[rgb]{0,0,1}$\circ$}.$\lambda_3$}                       \\
\multicolumn{1}{l}{{\color[rgb]{1,0,0}$\circ$}.$\lambda_1$}   & \multicolumn{1}{l}{{\color[rgb]{1,0,0}$\circ$}.$\lambda_1$}                       &  \multicolumn{1}{l}{{\color[rgb]{0,1,0}$\circ$}.$\lambda_2$}   & \multicolumn{1}{l}{{\color[rgb]{0,1,0}$\circ$}.$\lambda_2$}                       & \multicolumn{1}{l}{{\color[rgb]{0,0,1}$\circ$}.$\lambda_3$}   & \multicolumn{1}{l}{{\color[rgb]{0,0,1}$\circ$}.$\lambda_3$}                       \\ \hline 
\multicolumn{1}{l}{} &                       & \multicolumn{1}{l}{{\color[rgb]{0.4,0.4,0.4}$\circ$}} &                       \multicolumn{1}{l}{{\color[rgb]{0.4,0.4,0.4}$\circ$}} & \multicolumn{1}{l}{} &                       \\
\multicolumn{1}{l}{}   &                       & \multicolumn{1}{l}{{\color[rgb]{0.4,0.4,0.4}$\circ$}}   & \multicolumn{1}{l}{{\color[rgb]{0.4,0.4,0.4}$\circ$}}                       & \multicolumn{1}{l}{}   &                       \\ \hline 
                       & \multicolumn{1}{c}{}  & \multicolumn{2}{l}{{\color[rgb]{0.4,0.4,0.4}$\bullet$}}                           &                        & \multicolumn{1}{c}{}  \\
                       & \multicolumn{1}{c}{}  & \multicolumn{2}{l}{}                       &                        & \multicolumn{1}{c}{} \\ \hline 
\end{tabular}
\caption{Illustration of color transformation and downsampling used in \cite{FSIM, VSI, SCQI}. From first row to fourth row: four RGB pixels, RGB pixels multiplied by conversion coefficients, results of transformation into luminance channel, and downsampled results by a factor of 2. The same procedure should be repeated for the other two chromaticity channels ($C_1$ and $C_2$).}
\label{fig1}
\end{figure}

FSIM$_c$, VSI, and SCQI are FR-IQA metrics with state-of-the-art performance. Given the reference or distorted RGB input image, these metrics first convert it into a luminance ($L$) and two chromaticity channels ($C_1$ and $C_2$) with the following linear operators:

\begin{equation}
\small
  \ \begin{bmatrix}
L \\ C_1 \\ C_2 \end{bmatrix} = \begin{pmatrix}
\lambda_1 & \lambda_2 & \lambda_3 \\ \lambda_4 & \lambda_5 & \lambda_6 \\ \lambda_7 & \lambda_8 & \lambda_9 \end{pmatrix}  \begin{bmatrix} R \\ G \\ B \end{bmatrix}
  \label{operators}
\end{equation} 

where, values of $\lambda_{1...9}$ coefficients vary depending on the color space. The aforementioned metrics apply average filtering of size $M \times M$ on each converted channel, and downsample them by a factor of $M$. The value of $M$ is set to $[min(h, w)/256]$ \cite{Wangsite}, where $h$ and $w$ are image height and width, and $[.]$ is the round operator. The average filtering for luminance channel can be computed by the following equation:    

\begin{equation}
\small
  \ g_{ij} = \sum_{k=0}^{M-1} \sum_{l=0}^{M-1} w_{kl} ~ L_{i+k, j+l}
  \label{operators}
\end{equation} 

where, $\omega_{kl} = \frac{1}{M^2}$, and $g$ is the filtered image. In this paper, height and width of the filter are considered to be equal. Fig. \ref{fig1} illustrates this procedure for four RGB pixels converted into a typical channel ($L$), and downsampled by a factor of two ($M=2$). Exactly the same procedure should be repeated for the other two channels ($C_1$ and $C_2$).

\section{Alternative strategy} 

\begin{figure}[htb]
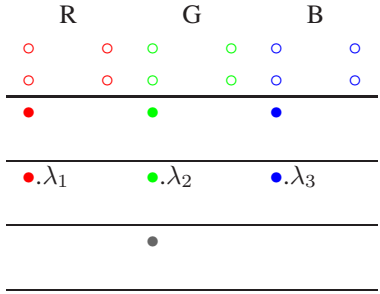

\centering
\begin{tabular}{llllll}
\multicolumn{2}{c}{R}                          & \multicolumn{2}{c}{G}                          & \multicolumn{2}{c}{B}                          \\
\color[rgb]{1,0,0}$\circ$                      & \multicolumn{1}{c}{\color[rgb]{1,0,0}$\circ$} & \color[rgb]{0,1,0}$\circ$                      & \multicolumn{1}{c}{\color[rgb]{0,1,0}$\circ$} & \color[rgb]{0,0,1}$\circ$                      & \multicolumn{1}{c}{\color[rgb]{0,0,1}$\circ$} \\
\color[rgb]{1,0,0}$\circ$                      & \multicolumn{1}{c}{\color[rgb]{1,0,0}$\circ$} & \color[rgb]{0,1,0}$\circ$                      & \multicolumn{1}{c}{\color[rgb]{0,1,0}$\circ$} & \color[rgb]{0,0,1}$\circ$ & \multicolumn{1}{c}{\color[rgb]{0,0,1}$\circ$} \\ \hline 
\multicolumn{1}{l}{\color[rgb]{1,0,0}$\bullet$}  &                       & \multicolumn{1}{l}{\color[rgb]{0,1,0}$\bullet$}  &                       & \multicolumn{1}{l}{\color[rgb]{0,0,1}$\bullet$}  &                       \\
\multicolumn{1}{l}{}   &                       & \multicolumn{1}{l}{}   &                       & \multicolumn{1}{l}{}   &                       \\ \hline 
\multicolumn{1}{l}{{\color[rgb]{1,0,0}$\bullet$}.$\lambda_1$} &                       & \multicolumn{1}{l}{{\color[rgb]{0,1,0}$\bullet$}.$\lambda_2$} &                       & \multicolumn{1}{l}{{\color[rgb]{0,0,1}$\bullet$}.$\lambda_3$} &                       \\
\multicolumn{1}{l}{}   &                       & \multicolumn{1}{l}{}   &                       & \multicolumn{1}{l}{}   &                       \\ \hline 
                       & \multicolumn{1}{c}{}  & \multicolumn{2}{l}{{\color[rgb]{0.4,0.4,0.4}$\bullet$}}                           &                        & \multicolumn{1}{c}{}  \\
                       & \multicolumn{1}{c}{}  & \multicolumn{2}{l}{}                       &                        & \multicolumn{1}{c}{} \\ \hline 
\end{tabular}
\caption{Illustration of the suggested downsampling and color transformation strategy. From first row to fourth row: four RGB pixels, downsampled results by a factor of 2 for each channel, downsampled results multiplied by conversion coefficients, results of transformation into luminance channel. For conversion to the chromaticity channels, e.g. $C_1$ and $C_2$, no downsampling is required.}
\label{fig2}
\end{figure}

We show that downsampling and color transformation is a more efficient strategy than color transformation and downsampling. Fig. \ref{fig2} illustrates how downsampling and color transformation is the better choice. In the suggested strategy, conversion to the other two channels (e.g. $C_1$ and $C_2$) is done using the already downsampled RGB channels. In fact, conversion is applied on the reduced size RGB channels instead of original size RGB channels. Note that both old strategy and suggested strategy need the same number of downsampling operations, but different conversion operations. The positive effect of the proposed strategy vary depending on the downsampling factor $M$.

\section{Experiments}

In experiments, three state-of-the-art FR-IQAs including FSIM$_c$ \cite{FSIM}, VSI \cite{VSI}, and SCQI \cite{SCQI} were chosen. Note that many other metrics in literature could have been used in this experiment. Table \ref{table1} lists the run times of three IQA models when applied on images of size 384x512 \cite{TID2013}, 1080x1920 \cite{ESPL}, and 2160x3840 \cite{4K}. The experiments were performed on a Corei7 3.40GHz CPU with 16 GB of RAM. The IQA models are tested in MATLAB 2013b running on Windows 7. The source codes of these metrics are provided by their authors \cite{FSIMweb, VSIweb, SCQIweb}. Only color space conversion and downsampling parts of these codes are replaced with the suggested strategy without optimizing other parts. It can be seen that these metrics run faster by using the suggested downsampling and conversion strategy. This improvement is more significant for larger images. More efficient indices are of high interest \cite{GMSD}, and can be used in real-time IQA applications.  

Another observation from the Table \ref{table1} is that the ranking of indices might not be the same when they are using different downsampling and conversion strategies or they are tested on images of different size. In previous works \cite{VSI, SCQI}, run times of metrics is compared on images with small size. However, images in real applications might be much larger.

\begin{table}[htb]
\centering
\scriptsize
\caption{Run time comparison of three state-of-the-art IQA models in terms of milliseconds. Run-times in bold indicate to the suggested downsampling and transformation strategy.}
{\label{table1}}
{\begin{tabular}{|c|c|c|c|c|c|c|}
\hline
IQA model & \multicolumn{2}{c|}{384x512} & \multicolumn{2}{c|}{1080x1920} & \multicolumn{2}{c|}{2160x3840} \\ \hline
FSIM$_c$ \cite{FSIM}     & 141.20   & \textbf{129.61}   & 605.91    & \textbf{499.68}    & 1556.86   & \textbf{1107.15}   \\ \hline
VSI \cite{VSI}       & 106.50   & \textbf{98.08}    & 498.28    & \textbf{414.57}    & 1973.08   & \textbf{1623.02}   \\ \hline
SCQI \cite{SCQI}      & 71.47    & \textbf{53.51}    & 512.04    & \textbf{318.97}    & 1918.80   & \textbf{1106.39}   \\ \hline
\end{tabular}}{}
\end{table}

The ranking of these metrics using their original conversion and downsampling strategy and the suggested strategy respectively is listed below.

\begin{enumerate}

\item 384x512: SCQI, VSI, FSIM$_c$ (no change)
\item 1080x1920: VSI, SCQI, FSIM$_c$ $\Rightarrow$ SCQI, VSI, FSIM$_c$
\item 2160x3840: FSIM$_c$, SCQI, VSI $\Rightarrow$ SCQI/FSIM$_c$, VSI

\end{enumerate}

It is worth to mention that there is no change in quality scores when old strategy is replaced with the suggested strategy. In other words, their quality prediction accuracy remain unchanged. The suggested conversion strategy should be used with caution. If a metric only uses a luminance channel \cite{SSIM, MSSSIM, GS}, the old strategy should be used.


\section{Conclusion}
\label{conclusion}

This paper suggests a more efficient strategy for those image quality assessment metrics that use downsampling and conversion to another color space. It was shown that downsampling first and then color space conversion is the better choice. In addition, run times of metrics should be compared on images with different size. The suggested strategy in this paper is highly recommended to be used in design of future IQA metrics in order to increase their efficiency, and to be able to compare their efficiency in a fair basis.

\section*{Acknowledgments}
The authors thank the NSERC of Canada for their financial support under Grants RGPDD 451272-13 and RGPIN 138344-14.


\bibliographystyle{IEEEtran}
\bibliography{egbib2}   

\end{document}